\documentclass[11pt,twoside]{article}
\usepackage{./asp2014}

\aspSuppressVolSlug
\resetcounters

\begin{document}

\title{Type II Cepheids as stellar tracers and distance indicators}
\author{Anupam Bhardwaj$^1$, Vittorio F. Braga$^{2,3}$, Dante Minniti$^{4,5,6}$, Rodrigo Contreras Ramos$^{4,7}$ and Marina Rejkuba$^8$}
\affil{$^1$Kavli Institute for Astronomy and Astrophysics, Peking University, Yi He Yuan Lu 5, Hai Dian District, Beijing 100871, China; \email{abhardwaj@pku.edu.cn}}
\affil{$^2$INAF - Osservatorio Astronomico di Roma, Italy}
\affil{$^3$ASI - Space Science Data Center, Italy}
\affil{$^4$Instituto Milenio de Astrof\'isica, Santiago, Chile}
\affil{$^5$Departamento de F\'isica, Facultad de Ciencias Exactas, Universidad Andr\'es Bello, Fern\'andez Concha 700, Las Condes, Santiago, Chile}
\affil{$^6$Vatican Observatory, V00120 Vatican City State, Italy}
\affil{$^7$Pontificia Universidad Cat\'olica de Chile, Instituto de Astrof\'isica, Av. Vicu\~na Mackenna 4860, 7820436, Macul, Santiago, Chile}
\affil{$^8$European Southern Observatory, Karl-Schwarzschild-Stra\ss e 2, 85748, Garching, Germany}

% This section is for ADS Processing.  There must be one line per author.
\paperauthor{Anupam Bhardwaj}{anupam.bhardwajj@gmail.com}{0000-0001-6147-3360}{Peking University}{Kavli Institute for Astronomy and Astrophysics}{Beijing}{China}{100871}{China}
%\paperauthor{Sample~Author2}{Author2Email@email.edu}{ORCID_Or_Blank}{Author2 Institution}{Author2 Department}{City}{State/Province}{Postal Code}{Country}

\begin{abstract}
Type II Cepheids are both useful distance indicators and tracers of old age stellar populations in their host galaxy. We summarize near-infrared observations of
type II Cepheids in the Large Magellanic Cloud and discuss the absolute calibration of their Period-Luminosity relations. Combining with the near-infrared data for type
II Cepheids in the Galactic bulge from the VISTA VVV survey, we estimated a robust distance to the Galactic center. We found that type II Cepheids trace the spherically symmetric 
spatial distribution with a possible evidence of ellipsoidal structure, similar to RR Lyrae stars. Together with {\it Gaia} and
VVV proper motions, type II Cepheids were found to trace the old, metal-poor, kinematically hot stellar populations in the Galactic bulge.
\end{abstract}

\section{Introduction}
Type II Cepheids (T2Cs) typically belong to low mass, old age ($>$10 Gyr) stellar populations found in the globular clusters, Galactic bulge and disc \citep{wallerstein2002}. 
T2Cs are generally classified in three subclasses based on their pulsation period: BL Herculis (BLH, $1\lesssim\!P\!\lesssim4$~d), W Virginis 
(WVI, $4\lesssim\!P\!\lesssim20$~d) and RV Tauri (RVT, $P\!\gtrsim 20$~d). These subclasses represent different evolutionary states going from post-horizontal branch to
asymptotic giant branch (AGB) and post AGB stars. A subclass of peculiar W Virginis (PWV) was suggested by \cite{soszynski2008a} for the 
most blue and bright WVI stars due to their distinct lightcurves.

T2Cs follow a Period-Luminosity relation (PLR) that is different from the bright classical Cepheids, which initially limited their use as distance indicators. 
However, these T2C variables are 1-3 mag brighter than the horizontal branch RR Lyrae stars and now widely used as distance tracers thanks to the increasing infrared observations in 
the past decade. The PLRs are intrinsically tight and less sensitive to metallicity and extinction at longer wavelengths \citep{matsunaga2006, matsunaga2011, ripepi2015, bhardwaj2017a, groenewegen2017, braga2018}. 

While Optical Gravitational Lensing Experiment \citep[OGLE,][]{soszynski2008a, soszynski2010, soszynski2017} survey has provided an extensive catalogue of 
T2Cs in the Magellanic Clouds and the Galactic bulge, their near-infrared counterparts are available from the surveys like VISTA survey of the Magellanic Clouds 
\citep[VMC,][]{cioni2011} and VISTA Variables in the V\'ia L\'actea \citep[VVV,][]{minnitivvv2010}.
These variables provide a unique opportunity to trace the structure and kinematics of underlying stellar populations in these galaxies, independent of other population II tracers such as RR Lyrae.  
For example, intermediate to old age metal rich red clump stars trace the X-shaped structure in the Galactic bulge \citep{mcwilliam2010, zoccali2016}. 
However, metal-poor red clumps and RR Lyrae stars trace a centrally concentrated spatial distribution, and also exhibit different kinematics \citep{valenti2016, zoccali2017}. 
Here, we summarize the results of T2C studies based on near-infrared surveys of the Large Magellanic Cloud (LMC) and the Galactic bulge. 

\vspace{-5pt}
\section{Period-Luminosity relations of Type II Cepheids in the LMC}
\vspace{-10pt}
\articlefigure[scale=0.32]{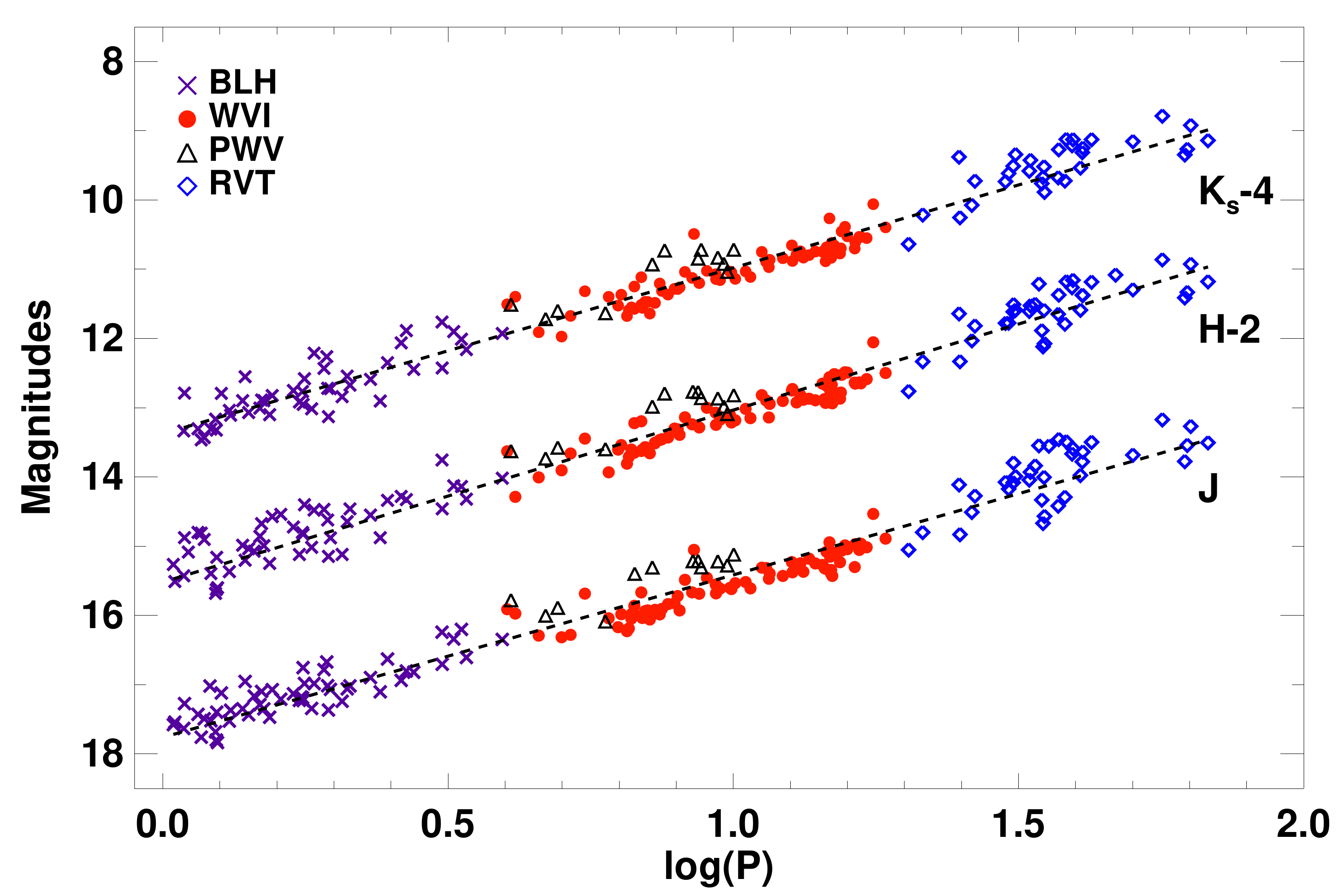}{fig:fig01}{Near-infrared PLRs for T2Cs in the LMC. The dashed line displays best fit linear regression over entire period range.}

\cite{macri2015} carried out a synoptic survey of the central 18 sq. deg. of the LMC at near-infrared wavelengths. This multi-epoch survey was aimed at providing absolute calibration of
classical Cepheid PLRs at $JHK_s$ wavelengths and resulted in an order of magnitude improvement in the precision of these relations \citep{macri2015, bhardwaj2016a, bhardwaj2016b}.
Note that the VMC survey \citep{cioni2011} has extended this work in $JK_s$-bands for almost all fields covered by the OGLE survey. \cite{bhardwaj2017a} crossmatched OGLE-III sample of
T2Cs with the synoptic survey of \cite{macri2015} and found 81 variables (16 BLH, 31 WVI, 12 PWVI, and 22 RVT) in common. The periods and classifications 
of these variables were adopted from the OGLE-III survey, and the new templates were used to fit the multi-epoch observations and estimate mean magnitudes in $JHK_s$ bands.

Having combined our sample with the data from \cite{matsunaga2011} and \cite{ripepi2015}, we show the near-infrared PLRs for T2Cs in the LMC in Figure~\ref{fig:fig01}. These PLRs
for BLH and WVI stars were found to be consistent with those in the globular clusters \citep{matsunaga2006}. The absolute calibration of T2C PLRs was done using the 
late type eclipsing binary distance to the LMC \citep{piet2013}. The trigonometric parallaxes for $kappa$ Pav and 5 RR Lyrae from {\it Hubble Space Telescope} were also used as 
additional zero point anchors to obtain robust distances to 26 Galactic globular clusters from \cite{matsunaga2006}. These distance estimates were found to be consistent with 
the measurements based on the empirical $M_V - [Fe/H]$ relation for horizontal branch stars.

\vspace{-5pt}
\section{Type II Cepheids in the Galactic bulge from the VVV survey}

The VVV survey is a multi-epoch near-infrared survey covering the Galactic bulge and the southern midplane \citep{minnitivvv2010}. \cite{bhardwaj2017b} crossmatched OGLE-III catalogue of bulge
T2Cs with the VVV survey and extracted a sample of $\sim340$ stars with well sampled $K_s$-band light curves. Assuming that the T2Cs in the bulge are located 
at the same distance, \cite{bhardwaj2017b} derived their PLRs and estimated a distance to the Galactic center, $R_0 = 8.34 \pm 0.03\textrm{(stat.)} \pm 0.41\textrm{(syst.)}$ kpc.
We used 3D extinction maps \citep{schul2014} based on the VVV survey data to iteratively estimate the extinction and individual distance estimates to T2Cs.
The spatial distribution of T2Cs displayed a centrally concentrated distribution, similar to RR Lyrae stars \citep{dekany2013}.

\citet{braga2018} extended the work on T2Cs in the VVV survey using a three times larger sample provided by the OGLE-IV catalogue. The left panel of 
Figure~\ref{fig:fig02} displays the PLR in $K_s$-band for T2Cs in the bulge. The absolute calibration of LMC PLRs is employed to estimate individual 
distances to T2Cs in the bulge. The mean distance to the Galactic center was found to be consistent with those based on stellar orbits around the 
central black hole \citep{gravity2019}. The T2C distribution shown in the right panel of Figure~\ref{fig:fig02} is very concentrated with a full width at
half maximum equivalent to 1.84 kpc. Note that no geometric corrections are applied to the histogram and the interested readers are referred to 
\citet{bhardwaj2017b} and \cite{braga2018} for details.

\articlefigure[scale=0.33]{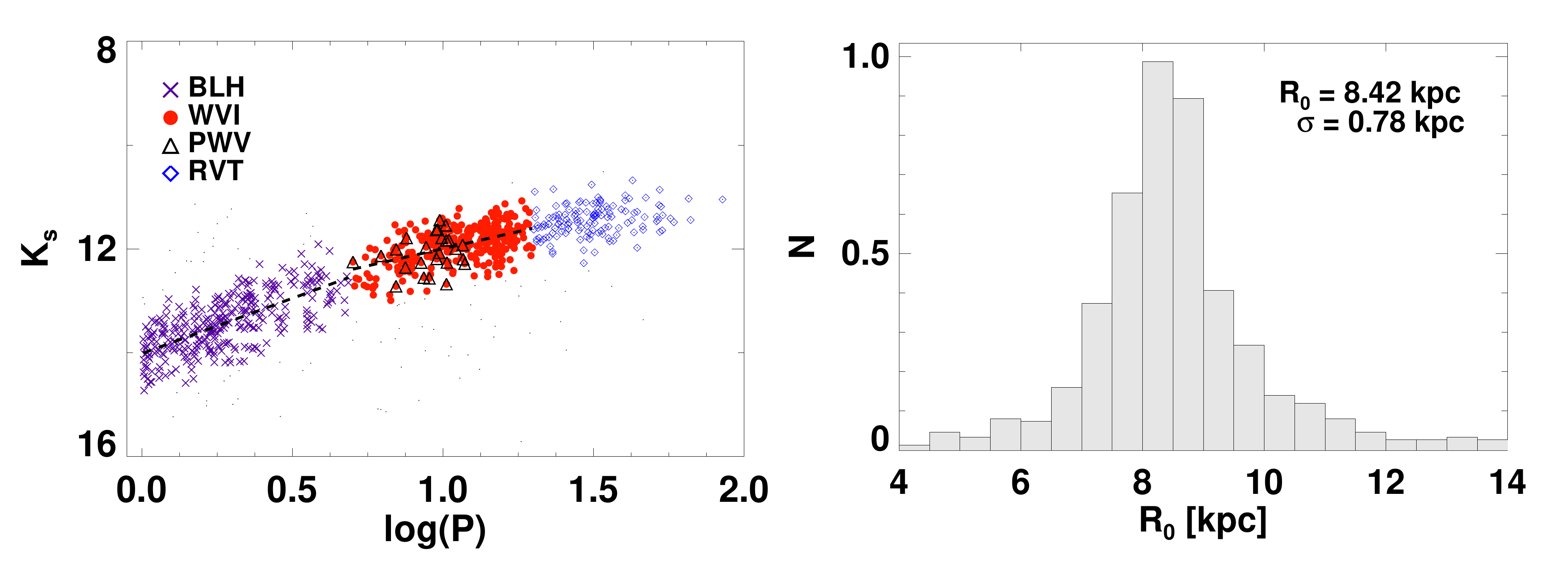}{fig:fig02}{{\it Left panel:} The PLR for T2C in the bulge. The dashed line represent two slope model fitted to BLH and WVR subclasses. Note 
that apparent brightness of RVT stars is close to the saturation limit of the VVV survey. {\it Right panel:} Histogram of individual distances to T2Cs anchored using the calibrated
LMC PLRs.}

\vspace{-5pt}
\section{Structure and Kinematics of Type II Cepheids in the Galactic bulge}
\cite{braga2018} also traced the spatial distribution of T2Cs and found a significant dependence of PLR on the galactic longitude and latitude. The T2Cs 
locate at $l>0\deg$  were found to be closer than those that are distributed at $l<0\deg$. The T2Cs display a centrally concentrated distribution 
(top left panel of Figure~\ref{fig:fig03}) and a possible evidence of ellipsoidal symmetry, similar to the RR Lyrae ellipsoid \citep{piet2015}.  Note that
the red clump giants overplotted on the top left panel of Figure~\ref{fig:fig03} trace the barred structure in the bulge. 

\articlefigure[scale=0.33]{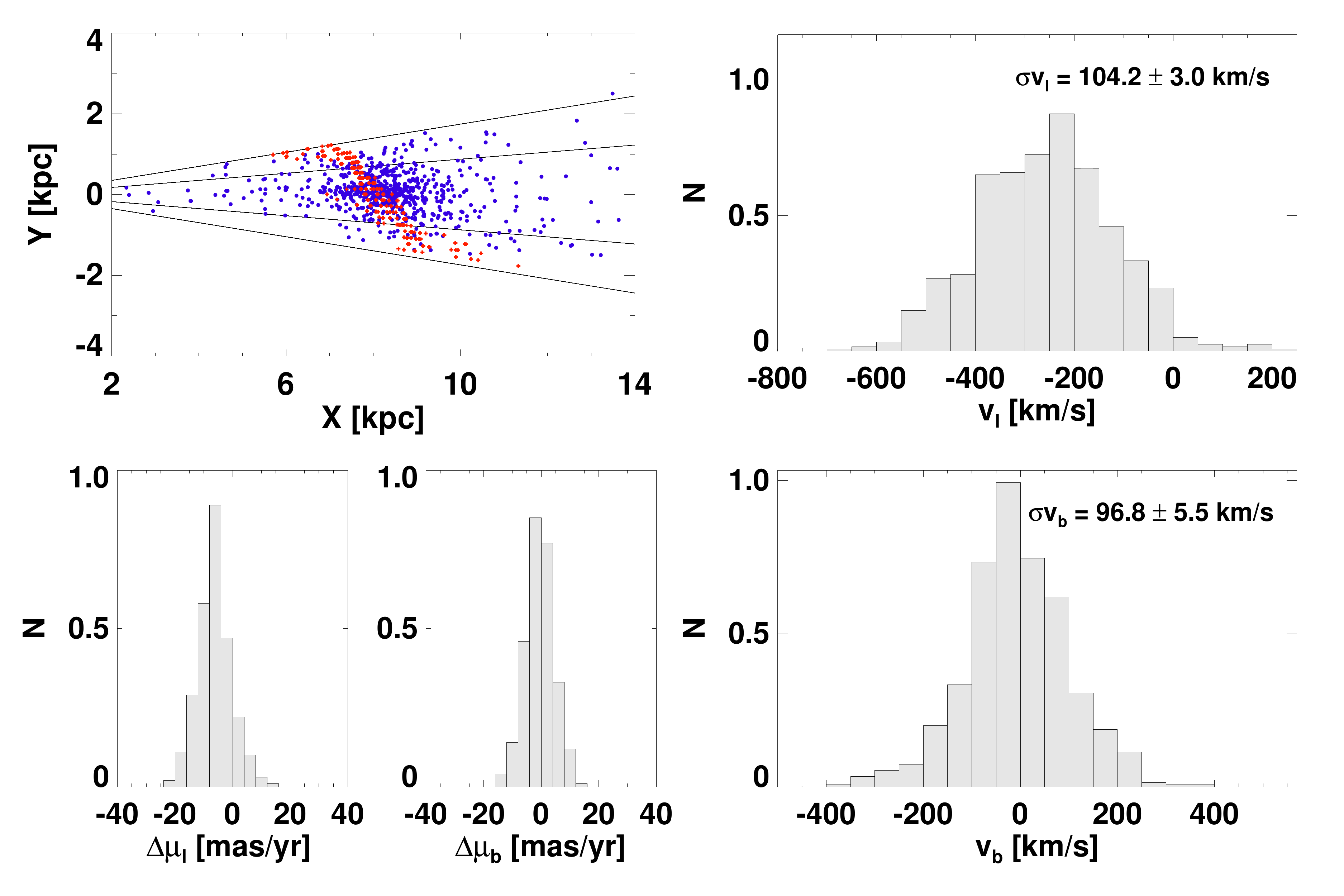}{fig:fig03}{{\it Top left:} Spatial distribution of T2Cs projected onto the Galactic plane. The solid lines represent the line of sight corresponding to $l = \pm 5\deg$ and $l \pm 10\deg$. The mean distances to red clump giants (red symbols) derived from the peaks of Gaussian distributions in each 1 sq. deg. fields are also shown. 
{\it Bottom left:} The proper motions of T2Cs located within central 2kpc in the bulge. {\it Top/bottom right:} Histograms of tangential velocity components of T2Cs in the bulge.}

The proper motions of T2Cs from {\it Gaia} and VVV were also used to investigate the kinematics of these stars in the bulge. Assuming that the center of mass of the old 
population overlaps with that of Sgr A$^\ast$ at the center of the Galaxy, proper motions of T2Cs can provide an indirect estimate of the center of mass.
The peaks of the proper motion distribution of T2Cs along the longitude and latitude directions are consistent with the proper motions of Sgr A$^\ast$  
(bottom left panel of Figure~\ref{fig:fig03}) Further, the velocity dispersions along both components were found to be fairly symmetric (see right panels of Figure~\ref{fig:fig03}) 
and the high velocity dispersions indicate that these stars belong to kinematically hot stellar populations in the Galactic bulge.

\vspace{-5pt}
\section{Conclusions}
We presented a summary of near-infrared observations of T2Cs in the LMC and derived their $JHK_s$-band Period-Luminosity relations. The absolute calibration of T2C
$K_s$-band Period-Luminosity relation in the LMC is used as an anchor to estimate the distance to the Galactic center based on the near-infrared data from the VVV survey. Individual distances to
T2Cs were used to trace their centrally concentrated spatial distribution in the bulge, which is similar to metal-poor red clumps and RR Lyrae stars. The proper motions of T2Cs
were used to trace their kinematics and it was found that these variables exhibit high velocity dispersions in the bulge. Thus T2C variables 
display structure and kinematic properties similar to those of other metal-poor, old stellar populations in the Milky Way bulge. 

\acknowledgements A.B. acknowledges research grant $\#$11850410434 awarded by the National Natural Science Foundation of China through the Research Fund for International Young Scientists,  China Postdoctoral General Grant (2018M640018), and Peking University Strategic Partnership Fund awarded to Peking-Tokyo Joint collaboration on Research in Astronomy and Astrophysics.

%\bibliography{../../mybib_final.bib}  % For BibTex
\end{document}